\documentclass[twocolumn,aps,floatfix,preprintnumbers]{revtex4}
\usepackage{psfig}
\begin{document}
\preprint{UH-511-1063-2004}
\preprint{LBNL-56686}
\title{QLC relation and neutrino mass hierarchy}
\author{Javier Ferrandis } 
\email{ferrandis @ mac.com}
\homepage{http: // homepage.mac.com / ferrandis}
\affiliation{ MEC postdoctoral fellow at the \\
 Theoretical Physics Group \\
 Lawrence Berkeley National Laboratory  \\
One Cyclotron Road, Berkeley CA 94720}
\author{Sandip Pakvasa } 
\email{pakvasa @ phys.hawaii.edu}
\affiliation{  Department of Physics \& Astronomy \\
 University of Hawaii at Manoa \\
 2505 Correa Road \\
 Honolulu, HI, 96822\\}
\begin{abstract}{
Latest measurements have revealed
that the deviation from a maximal 
solar mixing angle is approximately the Cabibbo angle ({\it i.e.} QLC relation). We argue that it is not
plausible that this deviation from maximality, be it a coincidence or not, 
comes from the charged lepton mixing.
Consequently we have calculated the required corrections
to the exactly bimaximal neutrino mass matrix ansatz 
necessary to account for the solar mass difference and the solar mixing angle.
We point out that the relative size of these two corrections depends strongly
on the hierarchy case under consideration.
We find that the inverted hierarchy case with opposite CP parities,
which is known to guarantee the RGE stability of the solar mixing angle,
offers the most plausible scenario for a high 
energy origin of a QLC-corrected bimaximal neutrino mass matrix. This possibility
may allow us to explain the QLC relation in connection with 
the origin of the charged fermion mass matrices.}
\end{abstract}
\maketitle
\newpage
%
\section{Introduction}
During the last year
our knowledge of the leptonic mixing matrix
has reached the precision level. 
The most recent $90$\% C.L. experimental 
results \cite{Ashie:2004mr,Araki:2004mb,Apollonio:2002gd}
and several global fits 
\cite{Bahcall:2004ut,Gonzalez-Garcia:2004it,Bandyopadhyay:2004da,Maltoni:2004ei}
have improved our knowledge of the neutrino mass differences and
indicate that the atmospheric mixing is almost maximal while
the solar mixing deviates from maximality in a particular way. In the standard notation,
\begin{eqnarray}
\sin \theta_{12} &=& 0.53 \pm 0.04   , \\
\sin \theta_{23} &=& 0.70\pm 0.11   , \\
\sin \theta_{13} &<& 0.15  ,\\
\Delta m^{2}_{\rm sun} &=& \Delta m^{2}_{\rm 21}  = (8.2 \pm 0.6)  \times 10^{-5} {\rm eV}^{2} , \\
\left| \Delta m^{2}_{\rm atm} \right| & =&  
\left|\Delta m^{2}_{\rm 32}\right|  = (2.45 \pm0.55)\times 10^{-3}{\rm eV}^{2} , 
\end{eqnarray}
We note that the mixing angle $\theta_{13}$ is constrained to be $\theta_{13}< 0.15$
by the non-observation of neutrino oscillations at the CHOOZ experiment \cite{Apollonio:2002gd}
and a fit to the global data \cite{Maltoni:2004ei}.   
This substantial improvement has confirmed that the leptonic mixing matrix, 
{\it heretoafter} called MNSP matrix 
\cite{MNSP},  is nearly bimaximal \cite{bimaximal,morebimax} and the
deviation from bimaximality observed has revealed a surprising relation 
between the Cabibbo angle, $\theta_{C}$ and the solar mixing angle \cite{Rodejohann:2003sc}, 
$$
\theta_{C} + \theta_{12} = 45.1^{\circ} \pm 2.4 ^{\circ}  \hbox{(1$\sigma$)},
$$
sometimes called the quark-lepton complementarity relation, 
{\it hereafter} referred to as QLC relation. There is a similar relation satisfied by the 
leptonic angle $\theta_{23}$ and the corresponding angle 
in the quark sector, although the errors are somewhat larger.  
Based on the experimental 
data it is convenient to define the following parametrization
\cite{Rodejohann1} of the mixing angles, 
\begin{eqnarray}
s _{23}  &=&  \frac{1}{\sqrt{2}} + \epsilon_{A} \lambda^{2}_{\nu}, \\
s_{12}   &=& \frac{1}{\sqrt{2}} \left( 1 - \lambda_{\nu} + \epsilon_{S} \lambda^{2}_{\nu} \right)  , \\
s_{13} &=&  \epsilon_{CP} \lambda^{2}_{\nu},
\end{eqnarray}
where $s _{ij} = \sin \theta_{ij}$ and 
the coefficients $\epsilon_{A}$, $\epsilon_{S}$ and $\epsilon_{CP}$ are
at most of order $\lesssim 4$, as indicated by the experimental uncertainities.
We note that we have defined the deviation from a maximal solar mixing angle as
$\lambda_{\nu}$ and not $\lambda=\theta_{C}$ 
to emphasize that $\lambda_{\nu}$ may not be exactly the Cabibbo angle. 
Therefore the MNSP matrix can be written to leading order in powers of $\lambda_{\nu}$ as, 
\begin{equation}
{\cal V}_{\rm MNSP}=
\left[
\begin{array}{ccc}
 \frac{1}{\sqrt{2}} \left( 1 + \lambda_{\nu} \right)  & - \frac{1}{\sqrt{2}} \left( 1 - \lambda_{\nu} \right) & 0 \\
 \frac{1}{2} \left( 1 - \lambda _{\nu}\right) &  \frac{1}{2} \left( 1 + \lambda_{\nu} \right) &  -\frac{1}{\sqrt{2}} \\ 
\frac{1}{2} \left( 1 - \lambda_{\nu} \right) &  \frac{1}{2} \left( 1 + \lambda_{\nu} \right) &   \frac{1}{\sqrt{2}} 
\end{array}
\right] + {\cal O}(\lambda_{\nu}^{2})
\end{equation}
The main implication of the QLC relation is fairly simple: the MNSP matrix
is to first order bimaximal \cite{bimaximal} and the deviation from the
exact bimaximality is a correction of the order of the Cabibbo angle,
{\it i.e.} around $20$\%. This resembles in certain way the 
situation in the quark sector, where it is known that to first order the CKM 
matrix is the unity matrix while the main correction is exactly the Cabibbo 
angle.

Explaining the QLC relation is a real challenge that any future theory of flavor
must address. Along with the extreme smallness of the neutrino masses,
this is another feature which qualitatively distinguishes the neutrino 
sector from the charged fermion sector.
The charged fermion spectra is very hierarchical, {\it i.e.}
the third generation masses are much heavier than the first and second generation fermion masses. 
Therefore we expect that there is a basis, probably the flavor basis 
(also known as lagrangian or symmetry basis), where 
the charged fermion diagonalization matrices are approximately diagonal.
On the other hand, it has been known for some time 
that the leptonic mixing matrix is nearly bimaximal.
It was expected that this distinctive feature could be explained
if the mechanism of neutrino mass generation is somehow disconnected from the
mechanism generating the flavor structure in the charged fermion sector. 
This may explain why many people, surprised by
the appearance of the Cabibbo angle in the leptonic mixing matrix,
have proposed to explain the QLC relation as a contamination coming from the
charged lepton mixing matrix.

In this paper we will analyze some generic implications of the QLC
relation for models of neutrino masses.
In Sec.~\ref{lepsec} we argue that it is 
not plausible that the QLC relation is explained by effects arising 
from the charged lepton mixing sector.
In Sec.~\ref{corrections} we analyze the form and relative size of the corrections to the 
bimaximal three neutrino mass matrix necessary to account for the QLC relation.
In Sec.~\ref{stability} we analyze the effects of the neutrino mass hierarchy on
the stability of the QLC relation and the implications for the scale of
neutrino mass generation.
In Sec.~\ref{RGEdsol} we analyze the possiblity that the solar mass
difference being zero at a high energy scale 
is RGE generated, triggered by a high energy origin
of the QLC relation. In Sec.~\ref{conclu} we summarize the main 
results of this paper.
\section{The QLC relation cannot arise from charged lepton mixing \label{lepsec}}
The MNSP mixing matrix is given by
\begin{equation}
{\cal V}_{\rm MNSP}=
({\cal V}^{l}_{L})^{\dagger} {\cal V}_{\nu}
\end{equation}
where ${\cal V}_{\nu}$ is the neutrino diagonalization matrix
and ${\cal V}^{l}_{L}$ is the left handed charged lepton diagonalization matrix,
${\cal M}_{l}^{\rm diag} = ({\cal V}^{l}_{L})^{\dagger}{\cal  M}_{l} {\cal V}_{R}^{l}$.
When trying to explain the QLC relation the first idea that comes to our mind
is the possibility that the QLC relation may arise from the charged lepton 
mixing matrix. We will argue that this is not plausible if one wants to understand
the well known empirical relations which connect the electron/muon mass ratio
with the quark sector. There is an empirical relation which has been known 
for quite a long time \cite{Cabibborelation,Georgi:1979df},
\begin{equation}
\left| V_{us} \right|
\approx
\left[ \frac{m_{d}}{m_{s}} \right] ^{\frac{1}{2}}
\approx 3 \left[ \frac{m_{e}}{m_{\mu}} \right]^{\frac{1}{2}},
\label{pattern1}
\end{equation}
This relation has been recently analyzed with precision by one of the authors
who noted that indeed the relation surprisingly works at the level of $\pm 16$\%, as the following
ratio shows (see Ref.~\cite{Ferrandis:2004ti} for details), 
\begin{eqnarray}
\left[ \frac{m_{d}}{m_{s}}\right] ^{1/2} :
\left[  \frac{m_{e}}{m_{\mu}} \right] ^{1/2}
&=& 3.06 \pm 0.48.
\label{Cabibbolepton} 
\end{eqnarray}
The relation between the Cabibbo angle and the down-strange quark mass ratio
can be simply explained, as known from the '70's\cite{Weinberg:hb},
if the down quark mass is generated from the mixing between
the first and second  families. Analogously, the relation between the Cabibbo angle 
and the electron-muon mass ratio can also be simply explained
if the electron mass is generated from the mixing between
the first and second lepton families. This implies that there is a leptonic
basis where the charged lepton mass matrix is given to leading order by,
\begin{equation}
\widehat{\cal M}_{l}=
\left[
\begin{array}{ccc}
0 & \left( \frac{m_{\mu} m_{e}}{m^{2}_{\tau}}  \right)^{\frac 1 2 }  &  {\cal O}(\lambda^{3})  \\
   \left( \frac{m_{\mu} m_{e}}{m^{2}_{\tau}}  \right)^{\frac 1 2 }    & 
  \left( \frac{m_{\mu}}{m_{\tau}} \right)  &  {\cal O}(\lambda^{2})  \\
  {\cal O}(\lambda^{3})   &  {\cal O}(\lambda^{2})  & 1 
\end{array}
\right].
\label{MLmatrix}
\end{equation}
Here $\lambda= \theta_{C}$.
The order of magnitude in the 
coefficients $(\widehat{\cal M}_{l})_{13}$ and 
$(\widehat{\cal M}_{l})_{23}$ can be obtained by requiring these entries 
not to affect the leading order terms for the charged lepton mass ratios.
From the matrix in Eq.~\ref{MLmatrix} and the empirical 
relation in Eq.~\ref{pattern1} it follows that the 
charged lepton mixing matrix to leading order is given in this leptonic basis by, 
\begin{equation}
{\cal V}_{L}^{l} \approx
\left[
\begin{array}{ccc}
1 & \lambda/3
& {\cal O}(\lambda^{3}) \\
\lambda/3
 & 1 & {\cal O}(\lambda^{2}) \\
{\cal O}(\lambda^{3}) & {\cal O}(\lambda^{2}) & 1 
\end{array}
\right]. 
\label{chlepmix}
\end{equation}
To sum up, Eq.~\ref{pattern1}
necessarily implies that there is a leptonic basis and a quark 
basis where the charged lepton mass matrix adopts the form given by Eq.~\ref{MLmatrix}
while the down-type quark mass matrix adopts a similar form
with $m_{\mu}/m_{\tau} = 3 m_{s}/m_{b}$.
It is very plausible that this is the flavor basis in some underlying 
theory of flavor.
For instance, 
this could be the basis where quarks and leptons unify in 
common representations of a Grand Unified group. 
It is known that some GUT models
can explain the relation in Eq.~\ref{pattern1} \cite{Georgi:1979df}.
This could be achieved if
the Higgs field giving mass to the charged leptons and down-type quarks
transforms under particular representations of the GUT group:
{\bf 45} in the SU(5) case or {\bf 126} in SO(10) models.
\begin{table*}
\begin{tabular}{|c|c|c|c|c|}
\hline
$
\begin{array}{c}
{\rm Normalized ~mass ~matrix} \\
\widehat{\cal M}_{\nu} 
\end{array} 
$
&
$
\begin{array}{c}
{\rm zero ~term} \\
\widehat{\cal M}_{\nu}^{\rm atm}
\end{array} 
$
& 
$
\begin{array}{c}
{\rm solar~ mass ~correction}\\
\widehat{\cal M}_{\nu}^{\rm sol}
\end{array} 
$
& 
$
\begin{array}{c}
{\rm QLC ~correction}\\
\widehat{\cal M}_{\nu}^{\rm QLC}
\end{array} 
$
&{\rm Eigenvalues} \\
\hline
\hline
{\rm normal~hierarchy}
& 
$
\left[
\begin{array}{ccc}
0 & 0 & 0 \\
0 & \frac{1}{2} & - \frac{1}{2} \\
0 & -\frac{1}{2} & \frac{1}{2} 
\end{array}
\right] 
$
&
$
\frac{\gamma }{2} 
\left[
\begin{array}{ccc}
1
& - \frac{1}{\sqrt{2}}  & - \frac{1}{\sqrt{2}} \\
- \frac{1}{\sqrt{2}}  & \frac{1}{2}  
& \frac{1}{2}   \\
-\frac{1}{\sqrt{2}} & \frac{1}{2}  
& \frac{1}{2} 
\end{array}
\right] 
$
&
$
\frac{\gamma }{2}
\left[
\begin{array}{ccc}
- 4  \lambda_{\nu}  & 0  & 0 \\
0 &  \lambda_{\nu}  &  \lambda_{\nu}   \\
0 &  \lambda_{\nu}   &  \lambda_{\nu}  
\end{array}
\right] 
$
& 
$
\begin{array}{c}
( 0 , \gamma  , 1)\\
\gamma \approx \lambda
\end{array}
$ 
\\ 
\hline
$\stackrel{\hbox{\rm inverted~hierarchy}}{\hbox{\rm with same CP parities}}$
&
$
\left[
\begin{array}{ccc}
1 & 0 & 0 \\
0 & \frac{1}{2}  & \frac{1}{2} \\
0 & \frac{1}{2} & \frac{1}{2}
\end{array}
\right] 
$
&
$
\frac{\gamma }{2}
\left[
\begin{array}{ccc}
1  & -\frac{1}{\sqrt{2}}  &  -\frac{1}{\sqrt{2}} \\
-\frac{1}{\sqrt{2}}& \frac{1}{2} & \frac{1}{2}  \\
 -\frac{1}{\sqrt{2}} & \frac{1}{2} 
& \frac{1}{2}  
\end{array}
\right] 
$
&  $
\frac{\gamma }{2}
\left[
\begin{array}{ccc}
- 2   \lambda_{\nu}  &  0 & 0 \\
0  &  \lambda_{\nu}  &  \lambda_{\nu}  \\
0 &   \lambda_{\nu} & \lambda_{\nu}
\end{array}
\right] 
$
 & 
 $
 \begin{array}{c}
 (1 , (1 +\gamma) , 0 )\\
 \gamma \approx \lambda^{2}/2
 \end{array}
$\\
\hline
$\stackrel{\hbox{\rm inverted~hierarchy}}{\hbox{\rm with opposite CP parities}}$
& 
$
\left[
\begin{array}{ccc}
0 & \frac{1}{\sqrt{2}} & \frac{1}{\sqrt{2}} \\
\frac{1}{\sqrt{2}} & 0 & 0 \\
\frac{1}{\sqrt{2}} & 0 & 0 
\end{array}
\right] 
$
&
 $
 \frac{\gamma }{\sqrt{2}}
\left[
\begin{array}{ccc}
-\frac{1}{\sqrt{2}}  & \frac{1}{2} &  \frac{1}{2} \\
 \frac{1}{2} & -\frac{1}{2\sqrt{2}}  & -\frac{1}{2\sqrt{2}}\\
 \frac{1}{2}  & -\frac{1}{2\sqrt{2}}  & -\frac{1}{2\sqrt{2}}  
\end{array}
\right] 
$
&
$
\left[
\begin{array}{ccc}
2  \lambda_{\nu} & 0 & 0 \\
0 & -   \lambda_{\nu}  & - \lambda_{\nu}  \\
0 & -   \lambda_{\nu}  & -  \lambda_{\nu}  
\end{array}
\right] $ 
 & 
 $
 \begin{array}{c}
 (1 , -(1 +\gamma) , 0 ) \\
 \gamma \approx \lambda^{2}/2
 \end{array}
 $\\
 \hline
\end{tabular} 
\caption{\rm Bimaximal zero order normalized neutrino mass 
matrices for the normal and inverted hierarchy cases
and their minimal first and second order corrections, which are 
necessary to account for the solar mass difference and the QLC relation.}
\label{zeroforms}    
\end{table*} 

It has been recently proposed\cite{chlepmixing,Minakata:2004xt} that,
to explain the deviation from a maximal solar mixing angle, one could assume that 
the neutrino mixing matrix in the flavor basis is exactly or approximately 
bimaximal, {\it i.e},
\begin{equation}
{\cal V}_{\nu} =  \left[
\begin{array}{ccc}
 \frac{1}{\sqrt{2}}  & - \frac{1}{\sqrt{2}}& 0 \\
 \frac{1}{2}&  \frac{1}{2} & - \frac{1}{\sqrt{2}} \\ 
\frac{1}{2} &  \frac{1}{2}  &   \frac{1}{\sqrt{2}} 
\end{array}
\right]. 
\end{equation}
and that the QLC relation is generated from charged lepton mixing.
We have pointed out above that most probably
the flavor basis of the underlying theory of flavor 
is the basis where quarks and leptons unify in common representations. In this basis 
we expect the charged lepton diagonalization matrix to be given by Eq.~\ref{chlepmix}.
Nevertheless, if this was the case we would obtain that 
$\theta_{12} = \frac{\pi}{4} + \frac{\theta_{C}}{6}$ instead of the observed QLC relation,
and this is quite inconsistent. 

If one insists to fully generate the observed deviation from 
bimaximality in the MNSP matrix from the charged lepton mixing,
assuming that the neutrino mixing matrix is approximately bimaximal in the
flavor basis, the required mixing in the charged lepton sector
would be very large and as a consequence 
the charged lepton mass matrix would adopt a very unnatural form
in the flavor basis in order to reproduce the correct electron mass 
\cite{Frampton:2004ud}. This kind of scenarios
do not provide a convincing explanation of the
precise relation that connects the charged lepton spectra and 
the quark spectra, see Eq.~\ref{pattern1}.

Therefore, most probably the bulk of the difference between $\theta_{12}$
and $\frac{\pi}{4}$ is already present in the neutrino mass matrix in the flavor basis,
or in other words the QLC relation must arise from the mechanism 
that generates the neutrino mass matrix and not
from the charged lepton mixing.
\section{QLC Corrected bimaximal mass matrices\label{corrections}}
The charged lepton mixing cannot account for the observed 
deviations from the bimaximal ansatz in the MNSP matrix.
Therefore, it is interesting to study the generic 
corrections to the bimaximal neutrino mass matrix 
that can account for the QLC relation. The form and relative size of these 
corrections can give us some insight in the origin of the neutrino mass matrix.
Let us denote the neutrino mass eigenstates by, 
\begin{equation}
{\cal M}_{\nu}^{\rm diag} = \left( m_{1}, m_{2}, m_{3} \right)
\end{equation}
Neglecting the charged lepton mixing, which can only give a second order contribution
to the QLC relation as we saw in the previous section,
the reconstructed neutrino mass matrix is, 
\begin{equation}
{\cal M}_{\nu} = {\cal V}_{\rm MNSP} {\cal M}_{\nu}^{\rm diag} {\cal V}_{\rm MNSP}^{\dagger}.
\end{equation}
This can be written as, 
\begin{equation}
{\cal M}_{\nu} =  {\cal M}^{\rm BiMax}_{\nu} + {\cal M}^{\rm QLC}_{\nu},
\end{equation}
where ${\cal M}^{\rm BiMax}_{\nu}$ is the well known bimaximal mass
matrix whose general expression is given by \cite{bimaximal}, 
\begin{equation}
{\cal M}^{\rm BiMax}_{\nu} =  \left[
\begin{array}{ccc}
\frac{1}{2} m_{12}  & \frac{1}{\sqrt{2}} \Delta_{12}& \frac{1}{\sqrt{2}} \Delta_{12} \\
\frac{1}{\sqrt{2}} \Delta_{12}&
\frac{1}{2} ( m_{12} +  m_{3} ) &
\frac{1}{2} ( m_{12}-  m_{3}  ) \\
\frac{1}{\sqrt{2}} \Delta_{12} &
\frac{1}{2} ( m_{12}-  m_{3} ) &
\frac{1}{2} ( m_{12} +  m_{3} ) 
\end{array}
\right]. 
\end{equation}
Here we have defined,
\begin{equation}
m_{12}= \frac{1}{2} (m_{1}+m_{2}) , \quad \Delta_{12}= \frac{1}{2  } (m_{1}-m_{2}). 
\end{equation}
The QLC correction, $\lambda_{\nu} = \pi/4 - \theta_{12}$, to the bimaximal ansatz
is generically given by,
\begin{equation}
{\cal M}^{\rm QLC}_{\nu} =  \left[
\begin{array}{ccc}
2   & 0&  0 \\
0 &
-1  &
-1\\
0 &
-1&
-1
\end{array}
\right] \lambda_{\nu} \Delta_{12}. 
\end{equation}
We note that we used $\lambda_{\nu}$ and not $\lambda=\theta_{C}$ 
to emphasize that $\lambda_{\nu}$ 
may not be exactly the Cabibbo angle.
Additionally the bimaximal mass matrix can be separated into two pieces, 
\begin{equation}
{\cal M}^{\rm BiMax}_{\nu} =  {\cal M}^{\rm atm}_{\nu} + {\cal M}^{\rm sol}_{\nu}
\end{equation}
The expressions 
for ${\cal M}^{\rm atm}_{\nu}$, ${\cal M}^{\rm sol}_{\nu}$ and ${\cal M}^{\rm QLC}_{\nu}$
depend on the hierarchy case under consideration. The particular forms
can be found in table~\ref{zeroforms}. Next we will comment on the main features
of the different hierarchy cases.
\subsection{Normal hierarchy case \label{seca}}
In the normal hierarchy case we obtain the leading order term
in the neutrino mass matrix assuming that
$m_{1}=0$ and $m_{2}=0$,
\begin{equation}
{\cal M}^{\rm atm}_{\nu}  = \frac{m_{3}}{2}
\left[
\begin{array}{ccc}
0 & 0 & 0 \\
0 & 1 & - 1 \\
0 & -1 & 1 
\end{array}
\right].
\end{equation}
This matrix generates mass for one neutrino,
$\nu_{3}$, which using the atmospheric mass difference, corresponds to,
\begin{equation}
m_{3} = \sqrt{ \Delta m^{2}_{\rm atm}} = (4.9 \pm 0.6)\times 10^{-2}~ \hbox{\rm eV}.
\end{equation}
To generate the solar mass difference we need to give mass 
to the neutrino $\nu_{2}$. To this end we need to introduce a small perturbation of the
previous matrix controlled by the parameter $\gamma = m_{2}/m_{3} \ll 1$.
To be consistent with bimaximal mixing we need the perturbation matrix to be of the form,
\begin{equation}
{\cal M}^{\rm sol}_{\nu}  = \gamma \frac{m_{3}}{2}
\left[
\begin{array}{ccc}
 1
& - \frac{1}{\sqrt{2}}  & - \frac{1}{\sqrt{2}} \\
- \frac{1}{\sqrt{2}}  & \frac{1}{2} 
& \frac{1}{2}  \\
-\frac{1}{\sqrt{2}} & \frac{1}{2} 
& \frac{1}{2} 
\end{array}
\right]. 
\label{msolnorm}
\end{equation}
In the normal hierarchy case $\gamma$ is related to the neutrino mass differences by,
\begin{equation}
\frac{(m^{2}_{2}- m^{2}_{1})}{(m^{2}_{3}-m^{2}_{2})} = \frac{ \gamma^{2}}{(1 - \gamma^{2})}
\approx \gamma^{2}. 
\end{equation}
Using experimental data $\gamma$ is determined to be,
\begin{equation}
\gamma \approx \left( \frac{\Delta m_{\rm sol}^{2}}{\Delta m_{\rm atm}^{2}} \right)^{\frac{1}{2}}
= 0.18 \pm 0.03.
\end{equation}
We note that $\gamma$ is curiously approximately the Cabibbo angle,
$\gamma \approx \lambda$,
this was noticed earlier in Ref.~\cite{Fishbane:1999yb}.
Finally to generate a deviation from maximality in the solar mixing angle
able to account for the QLC relation we need to introduce a second perturbation
given by,
\begin{equation}
{\cal M}^{\rm QLC}_{\nu}  =  \gamma \frac{m_{3}}{2} \lambda_{\nu}
\left[
\begin{array}{ccc}
- 4    & 0  & 0 \\
0 &   1  & 1  \\
0 &  1   &  1
\end{array}
\right] 
\label{mqlcnor}
\end{equation}
Therefore, in the normal hierarchy case, the correction to ${\cal M}^{\rm atm}_{\nu}$ 
coming from the matrix ${\cal M}^{\rm sol}_{\nu}$ is at most 
of order $\gamma \approx \lambda$, {\it i.e.} approx 20\%, 
in the entry (11) and approx. $ \lambda/2$ in the rest of entries of the matrix.
The entries in the QLC correction,  ${\cal M}^{\rm QLC}_{\nu}$, are
at most of order $4 \gamma \lambda_{\nu} \approx \lambda$ 
in the entry (11) and approx. $\lambda^{2}$ the rest.
Therefore for the normal hierarchy case to reproduce the neutrino data
we need the following hierarchy between the different corrections, 
\begin{equation}
  {\cal M}^{\rm QLC}_{\nu} \simeq {\cal M}^{\rm sol}_{\nu} <{\cal M}^{\rm atm}_{\nu} .
  \label{normhie}
\end{equation}
\subsection{Inverted hierarchy case with same CP parities \label{secb}}
In the inverted hierarchy case with same CP-parities
we obtain the leading order term 
in the neutrino mass matrix assuming that
$m_{1}=m_{2}$ and $m_{3}=0$,
\begin{equation}
{\cal M}^{\rm atm}_{\nu}  = m_{1}
\left[
\begin{array}{ccc}
1 & 0 & 0 \\
0 & \frac{1}{2} & \frac{1}{2}  \\
0 & \frac{1}{2}  & \frac{1}{2}  
\end{array}
\right].
\end{equation}
This matrix generates a degenerate mass for two neutrinos
which corresponds roughly to the atmospheric mass scale,
\begin{equation}
m_{1} = m_{2} = \sqrt{ \Delta m^{2}_{\rm atm}} 
\end{equation}
In this case, to generate the solar mass difference we need to break the 
degeneracy between the masses of $\nu_{1}$ and $\nu_{2}$. 
To this end we introduce a small perturbation of the form $m_{2} = m_{1} (1 + \gamma) $.
To be consistent with bimaximal mixing we need the perturbation matrix to be given by,
\begin{equation}
{\cal M}^{\rm sol}_{\nu}  = \frac{\gamma}{2} m_{1}
\left[
\begin{array}{ccc}
 1
& - \frac{1}{\sqrt{2}}  & - \frac{1}{\sqrt{2}} \\
- \frac{1}{\sqrt{2}}  & \frac{1}{2} 
& \frac{1}{2}  \\
-\frac{1}{\sqrt{2}} & \frac{1}{2} 
& \frac{1}{2} 
\end{array}
\right]. 
\label{msolinv}
\end{equation}
The solar mass difference is given by,
\begin{equation}
\Delta m^{2}_{\rm sol} =  (m^{2}_{2}- m^{2}_{1}) = m_{1}^{2} \gamma ( 2 + \gamma)
\approx 2  m_{1}^{2} \gamma.
\end{equation}
In this case, $\gamma $ can be determined from experimental data to be given by,
\begin{equation}
\gamma \approx  \frac{1}{2} \frac{\Delta m_{\rm sol}^{2}}{\Delta m_{\rm atm}^{2}} 
\approx \frac{1}{2} \lambda^{2} \approx 0.024
\end{equation}
Finally to generate a deviation from maximality in the solar mixing angle
able to account for the QLC relation we need to introduce a second perturbation
given by,
\begin{equation}
{\cal M}^{\rm QLC}_{\nu}  =  \gamma m_{1}  \lambda_{\nu}
\left[
\begin{array}{ccc}
-  1  & 0  & 0 \\
0 &  \frac{ 1 }{2} & \frac{1}{2}  \\
0 &  \frac{1}{2}   & \frac{ 1 }{2} 
\end{array}
\right] 
\label{mqlcinv}
\end{equation}
Therefore, in the inverted hierarchy case with same CP-parities, 
the correction to ${\cal M}^{\rm atm}_{\nu}$ 
coming from the matrix ${\cal M}^{\rm sol}_{\nu}$ is at most 
of order $\gamma/2 \approx \lambda^{3}$ in the entry (11) and $\approx \lambda^{3}/2$ the rest.
The entries in the QLC correction,  ${\cal M}^{\rm QLC}_{\nu}$, are
at most a correction 
of order $\lambda^{3}/2$ in the entry (11) and $\approx \lambda^{4}$ the rest.
Therefore for the inverted hierarchy case with same CP-parities 
to reproduce the neutrino data
we need the following hierarchy between the different corrections, 
\begin{equation}
{\cal M}^{\rm QLC}_{\nu} \lesssim {\cal M}^{\rm sol}_{\nu}  \ll {\cal M}^{\rm atm}_{\nu} .
\label{invsameCPhie}
\end{equation}
\subsection{Inverted hierarchy case with opposite CP parities}
In the inverted hierarchy case with opposite CP-parities
we obtain the leading order term 
in the neutrino mass matrix assuming that
$m_{2}=-m_{1}$ and $m_{3}=0$,
\begin{equation}
{\cal M}^{\rm atm}_{\nu}  = \frac{m_{1}}{\sqrt{2}}
\left[
\begin{array}{ccc}
0 & 1 & 1 \\
1 & 0 & 0  \\
1 & 0  & 0
\end{array}
\right].
\end{equation}
As in the same parities case
we need to break the 
degeneracy between the masses of $\nu_{1}$ and $\nu_{2}$
to generate the solar mass difference.
To this end we introduce a small perturbation of the form $m_{2} = - m_{1} (1 + \gamma) $.
To be consistent with bimaximal mixing we need the perturbation matrix to be given by,
\begin{equation}
{\cal M}^{\rm sol}_{\nu}  = \gamma \frac{m_{1}}{\sqrt{2}}
\left[
\begin{array}{ccc}
 - \frac{1}{\sqrt{2}}
& \frac{1}{2}  &  \frac{1}{2} \\
\frac{1}{2}  & -\frac{1}{2\sqrt{2}} 
& -\frac{1}{2\sqrt{2}}  \\
\frac{1}{2} & -\frac{1}{2\sqrt{2}} 
& -\frac{1}{2\sqrt{2}} 
\end{array}
\right]. 
\end{equation}
The solar mass difference is again given by,
\begin{equation}
\Delta m^{2}_{\rm sol} =  (m^{2}_{2}- m^{2}_{1}) = m_{1}^{2} \gamma ( 2 + \gamma).
\approx 2  m_{1}^{2} \gamma.
\end{equation}
Therefore $\gamma \approx \lambda^{2}/2$.
Finally to generate a deviation from maximality in the solar mixing angle
able to account for the QLC relation we need to introduce a
second perturbation given by,
\begin{equation}
{\cal M}^{\rm QLC}_{\nu}  =  \frac{m_{1}}{\sqrt{2}}  \lambda_{\nu}
\left[
\begin{array}{ccc}
 2\sqrt{2}  & 0  & 0 \\
0 & -\sqrt{2} & - \sqrt{2}\\
0 &  -\sqrt{2}  & -\sqrt{2}
\end{array}
\right]. 
\end{equation}
Therefore, in the inverted hierarchy case with same CP-parities, 
the correction to ${\cal M}^{\rm atm}_{\nu}$ 
coming from the matrix ${\cal M}^{\rm sol}_{\nu}$ is at most 
of order $\gamma/2\sqrt{2} \approx \lambda^{3}/\sqrt{2}$.
Interestingly the size of 
the entries to the QLC correction depends upon ${\rm sign} (m_{2})$
and in the opposite CP-parities case under consideration
we obtain that ${\cal M}^{\rm QLC}_{\nu}$ is between $\sqrt{2} \lambda_{\nu}$ and
$2 \sqrt{2} \lambda _{\nu}\approx 2/3$, {\it i.e.} approximately between 30\% and 60\%
of the leading term.
Therefore for the inverted hierarchy case with opposite CP-parities
to reproduce the neutrino data
we need the following characteristic hierarchy between 
the different corrections, 
\begin{equation}
{\cal M}^{\rm sol}_{\nu} \ll
{\cal M}^{\rm QLC}_{\nu}  \lesssim {\cal M}^{\rm atm}_{\nu} .
\end{equation}
This is very different from the hierarchies required for the corrections 
generated in the normal hierarchy case and inverted hierarchy case with same CP-parities.
In those two cases the QLC correction was of the same order or smaller than
the solar correction respectively.
\subsection{Generalization to the Dirac case \label{dirac}}
It is straightforward to extend the previous results to the case that
neutrinos are Dirac fermions. We will assume again that the mixing in the 
charged lepton sector in the flavor basis is very small, as a consequence
the MNSP matrix is very approximately the left-handed neutrino diagonalization matrix. We obtain, 
\begin{equation}
{\cal M}_{\nu} {\cal M}_{\nu}^{\dagger}
 = {\cal V}_{\rm MNSP} ({\cal M}_{\nu}^{\rm diag})^{2} {\cal V}_{\rm MNSP}^{\dagger}.
\end{equation}
We can generalize the results of Secs.~\ref{seca} and \ref{secb}
for the normal and inverted hierarchy cases.
In the first case we will introduce the same 
perturbation required to generate the solar mass difference, {\it i.e.} $m_{2} = \gamma m_{1}$.
The $({\cal M}_{\nu} {\cal M}_{\nu}^{\dagger})^{\rm sol}$
and $({\cal M}_{\nu} {\cal M}_{\nu}^{\dagger})^{\rm QLC}$ perturbations 
can be obtained from Eqs.~\ref{msolnorm} and \ref{mqlcnor} 
by implementing the substitution $m_{1} \rightarrow m_{1}^{2}$
and $\gamma \rightarrow \gamma^{2}$.
In the inverted hierarchy case we will now introduce the solar mass difference
perturbation in the form, $m_{2}^{2} = m_{1}^{2} (1 + \gamma^{2}) $.
In doing so we can obtain the perturbations
$({\cal M}_{\nu} {\cal M}_{\nu}^{\dagger})^{\rm sol}$
and $({\cal M}_{\nu} {\cal M}_{\nu}^{\dagger})^{\rm QLC}$ 
by implementing the same substitution,
$m_{1} \rightarrow m_{1}^{2}$ and $\gamma \rightarrow \gamma^{2}$,
in Eqs.~\ref{msolinv} and \ref{mqlcinv}.
Nevertheless, the perturbation parameter $\gamma$ will be determined in this case by
$\gamma^{2} \approx  \Delta m_{\rm sol}^{2}/\Delta m_{\rm atm}^{2}
\approx \lambda^{2} $. Therefore we will obtain for the normal and inverted hierarchy cases
corrections to the bimaximal ansatz similar to those in Eqs.~\ref{normhie} and \ref{invsameCPhie}
respectively.
\section{Radiative stability of the QLC relation\label{stability}}
It has been known for some time that 
the RGE effects can considerably affect the neutrino mixing angles
\cite{firstRGEeffects,morefirstRGEeffects}.
These effects can be especially important in the context of SUSY SO(10) models,
which are of especial interest for neutrino physics,
since in this case all the three third generation Yukawa couplings 
can be large \cite{Tanimoto:1995bf,Ellis:1999my}.
The RGE effects also depend crucially on the type of neutrino 
mass hierarchy under consideration \cite{habaRGEs,Mohapatra:2003tw}.

In the normal hierarchy case the RGE effects are known to be very small and as a consequence
they cannot account for a RGE generation of the QLC and or $\Delta m^{2}_{\rm sol}$
that, as we have seen in the previous section, must be of the same order of magnitude. 
Interestingly,  in the inverted hierarchy 
case the RGE evolution of the solar mixing depends crucially on
the neutrino CP-parities \cite{habaRGEs,moreRGEs}. The RGE equation for the solar mixing angle
in this case adopts a simple form, which is valid for small $\theta_{13}$,
as experiments indicate, given by \cite{Antusch:2003kp}, 
\begin{equation}
\frac{d \theta_{12}}{d t} = \frac{ C h^{2}_{\tau}}{8 \pi^{2}}  s_{12} c_{12} s_{23}^{2} 
\frac{\Delta m^{2}_{\rm atm}}{\Delta m^{2}_{\rm sol}} \left( 1 + \cos(\phi_{1}-\phi_{2})\right)
+ {\cal O}(\theta_{13}).
\end{equation}
Here $t = \ln (\mu/\mu_{0})$, $\mu$  
is the renormalization scale and $\phi_{1,2}$ are the neutrino CP-phases.
We will assume that an exactly 
bimaximal neutrino mass matrix is
generated at high energies, $s_{12}=c_{12}=s_{23}= 1/\sqrt{2}$,
and that the solar and atmospheric neutrino mass differences are phenomenologically
acceptable, {\it i.e.} that $\Delta m^{2}_{\rm sol} /\Delta m^{2}_{\rm atm}\approx \lambda^{2}$.
We obtain for the RGE generated shift in the solar mixing angle,
\begin{equation}
\Delta \theta_{12} 
\approx \frac{ C h^{2}_{\tau}}{32 \pi^{2}}  \frac{1}{\lambda^{2}} 
\left( 1 + \cos(\phi_{1}-\phi_{2})\right) \ln \left( \frac{\Lambda}{m_{Z}} \right).
\end{equation}
Here $\Delta \theta_{12} = \theta_{12}(\Lambda) - \theta_{12}(m_{Z})$.
In the SM $C=3/2$ and $h^{2}_{\tau}= m^{2}_{\tau} / m^{2}_{t} \approx 10^{-4}$ and 
assuming that $\Lambda = 10^{16}$~GeV 
we obtain for the radiatively generated $\Delta \theta_{12}$,
\begin{equation}
\left. \Delta \theta_{12} \right|_{\rm SM}^{\rm max}
\approx  3 \times 10^{-4} \left( 1 + \cos(\phi_{1}-\phi_{2})\right) 
\end{equation}
We note that to fit the experimental results we should obtain 
$\Delta \theta_{12}  \approx \lambda$.
It has already been pointed out \cite{Minakata:2004xt} that in the SM this 
correction is very small and it cannot be the source of the QLC
relation nor perturb a possible high energy origin of the QLC relation
irrespective of the neutrino CP-parities. 

In the MSSM the situation is more complicated. In this case $C=-1$ and 
$h^{2}_{\tau} \approx \tan^{2} \beta m^{2}_{\tau} / m^{2}_{t}$,
where $\tan\beta$ is the well known ratio of MSSM Higgs vacuum
expectation values. 
This is relevant in the case of SUSY SO(10) models which 
require a large $\tan\beta$. Assuming $\tan\beta = 50$ we obtain, 
\begin{equation}
\left. \Delta \theta_{12} \right|_{\rm MSSM}^{\rm max}
\approx  - \frac{1}{2} \left( 1 + \cos(\phi_{1}-\phi_{2})\right) 
\label{delta12mssm}
\end{equation}
This shows that
for the same CP-parities case the solar mixing angle would
be unstable under RGE corrections as is well known. We cannot 
generate radiatively the QLC relation because the MSSM correction
has a sign contrary to the required to fit the experimental data, $\Delta \theta_{12}\approx \lambda$.
On the other hand, Eq.~\ref{delta12mssm} shows that
the solar mixing angle in the case of
an inverted neutrino spectra with a maximal CP-parity phase
difference between the heaviest eigenvalues 
will be especially stable since in that case 
$ \cos (\phi_{1} -\phi_{2}) = -1$ and as a consequence
$d \theta_{12}/{d \mu} = 0$. We note that the term ${\cal O}(\theta_{13})$ which
has not been included in the RGE for $\theta_{12}$ 
also cancels for opposite CP-parities\cite{Antusch:2003kp}.
This opens the possibility that the QLC relation is generated
at a high energy scale, remaining stable all the way down
to the electroweak scale.
\section{A QLC triggered $\Delta m^{2}_{\rm sol}$? \label{RGEdsol}}
Let us assume that a QLC corrected bimaximal neutrino mass matrix
is generated at some high energy scale. We have seen in the previous
section that if there is an inverted neutrino hierarchy with opposite 
CP-parities, {\it i.e.} $m_{1}=-m_{2}$, the QLC relation will remain
stable under RGE evolution.
It is interesting to study if an initial high-energy deviation from 
maximality in the solar mixing, like the one given by the QLC relation, 
can trigger the generation 
of the correct solar mass difference radiatively through RGE running.
In some cases the solar mass difference, as 
pointed out some time ago \cite{CEIN},
could be fully generated by RGE corrections.
We will assume a limit case where at high energy 
$\theta_{13}$ and $\delta$, the Dirac CP-phase, are zero. 
The RGE for $\Delta m^{2}_{\rm sol}$ is given in this case by a simple expression
\cite{Antusch:2003kp},
\begin{equation}
8 \pi^{2}\frac{d \Delta m^{2}_{\rm sol}}{d t} = \alpha \Delta m^{2}_{\rm sol} - 
 C h^{2}_{\tau} 2 s_{23}^{2} ( m^{2}_{2} c_{12}^{2} - m_{1}^{2} s_{12}^{2})
+ {\cal O}(\theta_{13})
\end{equation}
Assuming that at high energies 
$\theta_{12}= \pi/4 - \lambda$ and $\theta_{23}=\pi/4$ we obtain for the 
radiatively generated solar mass difference,
\begin{equation}
8 \pi^{2}\frac{d \Delta m^{2}_{\rm sol}}{d t} = \alpha \Delta m^{2}_{\rm sol} - 
 2 C \lambda h^{2}_{\tau}   \Delta m^{2}_{\rm atm} .
 \end{equation}
This equation has a simple analytical solution. In the SM where 
$C=3/2$ we obtain,
\begin{equation}
 \left. \Delta m^{2}_{\rm sol} (\mu) \right|_{\rm SM} \approx 
 \frac{ 3 m^{2}_{\tau} \lambda}{\alpha m_{t}^{2}} \Delta m^{2}_{\rm atm}
(1-  e^{\frac{\alpha}{8\pi^{2}} ln \left(\frac{\mu}{\Lambda} \right)}) .
\end{equation}
Assuming that $\Lambda= 10^{16}$~GeV and $\mu=m_{Z}$ we obtain 
$ \left. \Delta m^{2}_{\rm sol} (m_{Z}) \right|_{\rm SM} 
\approx  2.8 \times 10^{-5} \Delta m^{2}_{\rm atm}$,
which is too small to account for the observed solar mass difference.
On the other hand in the MSSM $C=-1$ and we obtain,
\begin{equation}
 \left. \Delta m^{2}_{\rm sol} (\mu) \right|_{\rm MSSM} \approx 
  (-) \frac{ 2 t_{\beta}^{2} m^{2}_{\tau} \lambda}{ \alpha m_{t}^{2}} \Delta m^{2}_{\rm atm}
(1-  e^{\frac{\alpha}{8\pi^{2}} ln \left(\frac{\mu}{\Lambda} \right)})  .
\end{equation}
Assuming that $\Lambda= 10^{16}$~GeV, $\mu=m_{Z}$ 
and $\tan\beta$ is large, $\tan\beta=50$, we obtain 
$\left. \Delta m^{2}_{\rm sol} (m_{Z}) \right|_{\rm MSSM}
\approx - 2 \lambda^{2} \Delta m^{2}_{\rm atm}$. Therefore the 
radiatively generated $ \Delta m^{2}_{\rm sol} (m_{Z}) $ is of
the right magnitude but unfortunately of the wrong sign.
The experimental data requires that $ \left. \Delta m^{2}_{\rm sol} (m_{Z}) \right|
\approx  \lambda^{2} \Delta m^{2}_{\rm atm}$.
Therefore a RGE generation of $\left. \Delta m^{2}_{\rm sol} \right|_{\rm MSSM}$ triggered
by a very high energy generation of the QLC perturbed
bimaximal scenario, assuming and inverted hierarchy with opposite CP-parities,
does not seem to be in agreement with the data.
\section{Conclusions \label{conclu}}
We have studied several model independent 
implications of the measured deviation from maximality
in the solar mixing angle. We have pointed out that it is not plausible
that this deviation is generated in the charged lepton mixing matrix.
We have studied the generic low energy corrections to the exactly bimaximal ansatz 
necessary to account for both the solar mass difference and a
non-maximal solar mixing angle.
We pointed out that the relative size of these corrections depends strongly on
the neutrino hierarchy under consideration.
For the normal and inverted hierarchy with same CP parities
it seems very difficult to understand the origin of the QLC relation independently 
from the origin of $\Delta m^{2}_{\rm sol}$ since the respective corrections
are of the same order of magnitude.
In that case the QLC relation is most probably a coincidence
unless the neutrino mass matrix is generated at low energy scales.

On the other hand,
for an inverted hierarchy with opposite CP parities
the correction to the bimaximal
ansatz necessary to explain the QLC relation is of the same order but
smaller than the leading term of the bimaximal matrix
and both are much larger than the correction necessary to generate $\Delta m^{2}_{\rm sol}$.
Additionally the leading bimaximal term as well as 
the QLC perturbation could both have a high energy origin since
the solar mixing angle is very stable under RGE effects.
This raises the possibility to link the origin of the QLC relation
with the origin of the charged fermion mass matrices. 
Although this setup does not allow us to radiatively generate
$\Delta m^{2}_{\rm sol}$ entirely by RGE corrections
there are other possible explanations available in the literature for
the origin of the measured $\Delta m^{2}_{\rm sol}$. 
We believe, as our analysis indicates, that 
the inverted hierarchy case with opposite CP-parities may be 
the most interesting possibility from a model building point of view
when searching for a non-coincidental, high-energy 
explanation of the QLC relation.
\acknowledgements
This work is supported by:
the Ministry of Science of Spain under grant EX2004-0238,
the US Department of Energy under Contracts: DE-AC03-76SF00098, 
DE-FG03-91ER-40676 and DE-FG03-94ER40833 and
by the National Science Foundation under grant PHY-0098840.

\end{document}